\documentclass[12pt,a4paper]{article}
\usepackage{amsfonts}
\usepackage{amsmath}
\usepackage{amssymb}
\usepackage{latexsym}
\numberwithin{equation}{section}

\newcommand{\lanln}[1]{$\langle$\texttt{arXiv:#1}$\rangle$}
\newcommand{\SLtwor}{{\mathrm{SL}}(2,\mathbb{R})}

\newcommand{\BbbR}{\mathbb{R}}
\newcommand{\BbbZ}{\mathbb{Z}}
\newcommand{\BbbC}{\mathbb{C}}
\newcommand{\be}{\begin{eqnarray}}
\newcommand{\ee}{\end{eqnarray}}

\newtheorem{proposition}{Proposition}[section]

\newcommand{\myproof}{\emph{Proof\/}. }

\newcommand{\Aobs}{{\mathcal{A}_\mathrm{obs}}}

\newcommand{\Haux}{{\mathcal{H}_\mathrm{aux}}}

\newcommand{\Hraq}{{\mathcal{H}_\mathrm{RAQ}}}

\newcommand{\Gammared}{{\Gamma_\mathrm{red}}}

\newcommand{\Omegared}{{\Omega_\mathrm{red}}}

\newcommand{\barGamma}{{\overline\Gamma}}

\title{Superselection sectors in the
Ashtekar-Horowitz-Boulware model}

\author{Jorma Louko\thanks{jorma.louko@nottingham.ac.uk}
\ and
Alberto Molgado\thanks{pmxam@nottingham.ac.uk}
\\
\noalign{\vspace{3ex}}
\small{\it School of Mathematical Sciences,
University of Nottingham,}\\
\small{\it Nottingham NG7 2RD, UK}
\\
\noalign{\vspace{1ex}}\\
\small{(Revised August 2005)}
\\
\noalign{\vspace{1ex}}
\small{\lanln{gr-qc/0505097}}
\\
\noalign{\vspace{2ex}}
\small{Published in {\it Class.\ Quantum Grav.\ \bf 22} 
(2005) 4007-4019}
}

\date{}

\begin{document}


\maketitle

\begin{abstract}
We investigate refined algebraic quantisation of the constrained
Hamiltonian system introduced by Boulware as a simplified version of
the Ashtekar-Horowitz model. The dimension of the physical Hilbert
space is finite and asymptotes in the semiclassical limit to
$(2\pi\hbar)^{-1}$ times the volume of the reduced phase space. The
representation of the physical observable algebra is irreducible for
generic potentials but decomposes into irreducible subrepresentations
for certain special potentials. The superselection
sectors are related to singularities in the reduced phase space and to
the rate of divergence in the formal group averaging integral. There
is no tunnelling into the classically forbidden region of the
unreduced configuration space, but there can be tunnelling between
disconnected components of the classically allowed region.
\end{abstract}

\newpage

\section{Introduction}

In this paper we study quantisation of the constrained Hamiltonian
system introduced by Boulware \cite{Boul} as a simplified version of
the Ashtekar-Horowitz model~\cite{AH}. Both systems have a
four-dimensional unreduced phase space and a single constraint,
quadratic in the momenta. The Ashtekar-Horowitz model was originally
introduced in \cite{AH} to model the situation occurring in general
relativity in which certain parts of the unreduced configuration space
are not in the projection of the constraint hypersurface. These parts
of the unreduced configuration space thus play no part in the
classical theory, but they could give rise to tunnelling effects in
Dirac-style quantisations
\cite{Dir,Hen}. 
The quantisation discussed in
\cite{AH} indeed displayed such effects, containing physical states
that have support in the classically forbidden region of the
configuration space. Further developments using a variety of
quantisation schemes can be found in \cite{Boul,Gotay, Tate, Ash2}. In
particular, there exists a path-integral
quantisation of Boulware's system that exhibits no tunnelling into
the classically forbidden region~\cite{Boul}.

We shall investigate Boulware's system within the refined algebraic
quantisation (RAQ) programme of~\cite{epistle,GM1,GM2} (for reviews,
see \cite{Giulini-rev,Marolf-MG}). The main new issue of interest for
us is that a RAQ quantum theory entails not just a physical Hilbert
space $\Hraq$ but also a precisely-defined algebra $\Aobs$ of physical
observables. We wish to study this algebra and in particular ask
whether its representation contains superselection sectors.

A major piece of technical input in RAQ is the rigging map, which maps
a dense subspace of suitably well-behaved states in the unconstrained
Hilbert space to distributional states that solve the constraints. In
our system the integral of matrix elements over the gauge group does
not converge in absolute value, which complicates attempts to define a
rigging map by group averaging. However, for generic potentials (in a
sense that will be made precise) the formal group averaging expression
nevertheless suggests a rigging map candidate: We show that this
candidate is a genuine rigging map and the resulting representation of
$\Aobs$ is irreducible. For certain special potentials the rigging map
candidate becomes ill defined, owing to formally divergent terms, but
we show that the candidate can then be replaced by a genuine rigging
map by renormalising the divergences. In this case the representation
of $\Aobs$ decomposes into superselection sectors, labelled by the
degrees of divergence in the formal rigging map candidate, and the
representation within each superselection sector is irreducible.
These results bear a qualitative similarity to the superselection
sector results found in
\cite{GoMa} in an $\mathrm{SO}(n,1)$ gauge system but our sense of
convergence in the group averaging is weaker and our superselection
sector structure is richer.

The system also exhibits a striking connection between quantum
superselection and classical singularities: Superselection sectors
exist precisely when some vectors in $\Hraq$ are supported on the part
of the unreduced configuration space that is associated with singular
parts of the reduced phase space.

The physical Hilbert space is finite dimensional, and in the
semiclassical limit its dimension asymptotes to $(2\pi\hbar)^{-1}$
times the volume of the reduced phase space. The only superselection
sector that remains significant in the semiclassical limit is the one
whose rigging map requires no renormalisation.

As in~\cite{Boul}, there is no tunnelling of the kind found in
$\cite{AH}$ into the classically forbidden region of the unreduced
configuration space. If the classically allowed region of the
unreduced configuration space is not connected, there can be
tunnelling between its components.

The rest of the paper is as follows. Section \ref{sec:classical}
introduces the classical system, and the quantisation is carried out
in section~\ref{sec:RAQ}. Section~\ref{sec:discussion} presents brief
concluding remarks. The proofs of certain technical results are
deferred to three appendices. We set $\hbar=1$ except in the
semiclassical limit discussion in section~\ref{sec:RAQ}.

\section{Classical system}
\label{sec:classical}

The configuration space of the system is $\mathcal{C} := T^2 \simeq
S^1 \times S^1$. We write the points in $\mathcal{C}$
as $(x,y)$, where $x\in S^1$ and $y\in S^1$, 
and points in the 
phase space $\Gamma := T^*\mathcal{C}$ 
as $(x,y,p_x,p_y)$,
where $p_x\in\BbbR$ and $p_y\in\BbbR$.

The action reads 
\be
S=\int\! dt\, \bigl(  p_x \dot{x} + p_y \dot{y} 
- \lambda C \bigr)
\ ,
\label{eq:hamilt}
\ee
where the overdot denotes differentiation
with respect to the parameter $t$ and  
$\lambda$ is a Lagrange multiplier associated with 
the constraint
\be
C := p_x^2 - R(y) \ ,
\label{eq:class-constraint}
\ee
where $R: S^1 \to \BbbR$ is smooth. 
We assume $R$ to be
positive at least somewhere. We also assume that $R$ has at most 
finitely many
stationary points. It follows that $R$ has at least two
stationary points and at most finitely many zeroes. We further assume
that each stationary point of $R$ has a
nonvanishing derivative of $R$ of some order. To simplify
the discussion of the classical system, we assume that no zero of $R$
is a stationary point. 
In section \ref{sec:RAQ} we will 
introduce a further genericity condition on $R$ 
to control the quantum theory. 

The constraint surface $\barGamma$ is the subset of $\Gamma$ where
$C=0$. By our assumptions about~$R$, $\barGamma$ is the Cartesian
product of $S^1 \times \BbbR = \{(x,p_y)\}$ with finitely many
disjoint circles in $S^1 \times \BbbR = \{(y,p_x)\}$. 
We show in appendix
\ref{app:Gammared} that each connected component of the reduced phase
space $\Gammared$ is a two-dimensional symplectic manifold with
certain one-dimensional singular subsets, and the symplectic volume of
$\Gammared$ is finite and equal to $2\pi \int_{R>0} |R'(y)| /
\sqrt{R(y)} \, dy$, or $4\pi$ times the total variation of $\sqrt{R}$
over the subset of $S^1$ on which $R$ is positive. The singularities
occur at the stationary points of~$R$: This will become important on
comparison to the quantum theory.

\section{Quantisation}
\label{sec:RAQ}

In this section we quantise the system, 
following refined algebraic quantisation 
as reviewed in~\cite{Marolf-MG}. 
Subsection 
\ref{Auxspace} fixes the structure in the auxiliary Hilbert space. 
The rigging map is constructed in subsection \ref{sec:p=1} 
under a certain genericity condition on $R$ and in subsection 
\ref{sec:p>1} under a weaker form of this condition.

\subsection{Auxiliary structure}
\label{Auxspace}

Our auxiliary Hilbert space $\Haux$ is the space of 
square integrable functions on $\mathcal{C}$ in the inner product 
\be
\left( \phi_1 ,\phi_2 \right)_{\mathrm{aux}}
:=
\int \! dx \, dy\, \overline{\phi_1
(x,y)}\phi_2 (x,y)
\ ,
\ee
where the overline denotes complex conjugation. To promote the
classical constraint (\ref{eq:class-constraint}) into a quantum
operator, we replace the momentum term $p_x^2$ by 
$- \partial^2/\partial x^2$, and in the
potential term we replace the function $R(y)$ 
by the operator ${\hat R}$ that acts on
$\phi\in\Haux$ by $\bigl({\hat R}\phi\bigr)(x,y) = R(y)\phi(x,y)$. The
quantum constraint $\hat{C}$ thus reads 
\be
\hat{C}:= - \frac{\partial^2}{\partial x^2} - \hat{R}
\ .
\ee
$\hat{C}$ is essentially self-adjoint on $\Haux$
and exponentiates into the one-parameter family of unitary operators 
\be
U(t):=e^{-it\hat{C}}
\ , 
\quad
t\in\BbbR
\ . 
\ee

We next need to choose the test space~$\Phi$, a linear subspace of
sufficiently well-behaved states in~$\Haux$. Taking advantage of the
Fourier decomposition in~$x$, we take $\Phi$ to be the space of
functions $f:\mathcal{C} \to \BbbC$ of the form $f(x,y) =
\sum_{m\in\BbbZ} e^{imx} f_m(y)$, where each $f_m: S^1 \to \BbbC$ is
smooth and only finitely many $f_m$ are different from zero for
each~$f$. $\Phi$~is clearly a dense linear subspace of~$\Haux$. If
$f\in\Phi$, then
\be
\bigl( U(t)f \bigr) (x,y) 
= 
\sum_m e^{-it [m^2-R(y) ]} e^{imx} f_m(y)
\ , 
\label{eq:U(t)}
\ee
which shows that $U(t) f \in \Phi$. $\Phi$~is thus invariant
under~$U(t)$. 
Note that if $f,\,g \in \Phi$, then
\be
{( f, g )}_{\mathrm{aux}} =
2\pi \sum_m \int \! dy\, \overline{f_m(y)} g_m(y)
\ . 
\ee

The above structure determines the RAQ observable algebra $\Aobs$ as
the algebra of operators ${\mathcal{O}}$ on $\Haux$ such that the
domains of ${\mathcal{O}}$ and ${\mathcal{O}}^\dag$ include~$\Phi$,
${\mathcal{O}}$ and ${\mathcal{O}}^\dag$ map $\Phi$ to itself
and
${\mathcal{O}}$ commutes with $U(t)$ on $\Phi$ for all~$t$. Note that
if ${\mathcal{O}}\in\Aobs$, then also ${\mathcal{O}}^\dag\in\Aobs$.

What remains is to specify the final ingredient in RAQ, an antilinear
rigging map 
$\eta: \Phi \to \Phi^*$, where the star denotes the algebraic
dual, topologised by pointwise convergence. 
$\eta$~must be real and positive, states in its image must
be invariant under the dual action of~$U(t)$, and $\eta$ must
intertwine with the representations of $\Aobs$ on $\Phi$ and $\Phi^*$
in the sense that for all ${\cal O}\in\Aobs$ and $\phi\in\Phi$,
\be
\eta ({\mathcal{O}} \phi) = {\mathcal{O}} ( \eta \phi ) 
\ . 
\label{eq:eta-intertwining-def}
\ee
In terms of the matrix elements, 
(\ref{eq:eta-intertwining-def}) reads 
\be
\eta({\mathcal{O}} \phi_1) [\phi_2]
= 
\eta(\phi_1) [{\mathcal{O}}^\dag \phi_2]
\ , 
\label{eq:eta-intertwining-matrix}
\ee
where $\phi_1, \, \phi_2 \in \Phi$, the left-hand side denotes the
dual action of $\eta({\mathcal{O}} \phi_1)\in\Phi^*$ 
on $\phi_2\in\Phi$ and 
the right-hand side denotes the dual action of 
$\eta(\phi_1) \in\Phi^*$
on ${\mathcal{O}}^\dag \phi_2 \in \Phi$. 
The rigging map then completely 
determines both the physical Hilbert space $\Hraq$
and the representation of $\Aobs$ on it: 
$\Hraq$ is the Cauchy completion of the image of
$\eta$ in the inner product
\begin{equation}
\bigl(\eta(\phi_1), \eta(\phi_2) \bigr)_{\mathrm{RAQ}} 
:= \eta(\phi_2)[\phi_1]
\ , 
\label{phys-ip}
\end{equation}
and the properties of $\eta$ and $\Aobs$ imply that $\eta$ induces an
antilinear representation of $\Aobs$ on~$\Hraq$, with the image of
$\eta$ as the dense domain. 

To find a rigging map, we shall place a genericity condition
on~$R$. In subsection \ref{sec:p=1} we work under a genericity
condition that is relatively strong and will make the representation
of $\Aobs$ irreducible. In subsection \ref{sec:p>1} we weaken this
condition in a way that will lead to superselection sectors.

\subsection{Rigging map for generic $R$}
\label{sec:p=1}

Our construction of the rigging map will use the solutions to the
equation
\be
q^2 = R(y) 
\ , 
\label{eq:q2=R}
\ee 
where the non-negative integer $q$ is a parameter and $y$ is regarded
as the unknown. We assume that (\ref{eq:q2=R}) has solutions for
some~$q$. From the assumptions on $R$ it follows that solutions only
exist for finitely many $q$ and that for each $q$ there are at most
finitely many solutions.

In this subsection we assume that none of the solutions to
(\ref{eq:q2=R}) are stationary points of~$R$. 
We write the
solutions as~$y_{qj}$, where the second index labels the solutions for
given~$q$. 

Recall that the group averaging proposal seeks a rigging map 
as an implementation of the formal expression 
\be
\eta: \phi \mapsto 
\int_{-\infty}^{\infty} \! 
dt \, \phi^\dag U(t) 
\ . 
\label{eq:eta-formal}
\ee
A~strategy proposed in \cite{GM2}
would be to try to define (\ref{eq:eta-formal}) in
terms of integrated matrix elements as 
\be
\eta(\phi_1)[\phi_2]
= 
\int_{-\infty}^{\infty} \! 
dt \, 
\bigl( \phi_1 , U(t) \phi_2 \bigr)_{\mathrm{aux}}
\ . 
\label{eq:eta-ga}
\ee
Examples in which this strategy can be successfully implemented are
found in~\cite{GoMa,LouRov,LouMol1,LouMol2}. In our system, however, a
saddle-point estimate shows that there are states for which the
absolute value of the integrand on the right-hand side of
(\ref{eq:eta-ga}) is asymptotically proportional to ${|t|}^{-1/2}$ as
$|t| \to \infty$, and for such states the integral is not absolutely
convergent. 
While it may be possible to work with (\ref{eq:eta-ga}) in some
appropriate weaker sense of conditional convergence, we shall not
pursue this line here. Instead, we show that a formal reinterpretation
of (\ref{eq:eta-ga}) leads to a map that can be
directly proven to be a rigging map. 

The integral expression (\ref{eq:eta-formal}) can be formally
rewritten as (cf.\ \cite{epistle, QORD,BC}) 
\be
\bigl(\eta (f) \bigr) (x,y)
=  
2\pi\, \sum_m 
\delta \bigl(m^2-R(y)\bigr) \, 
e^{-imx} \overline{f_m (y)}
\ , 
\label{eq:eta-ftilde1-int}
\ee
or equivalently as 
\be
\bigl(\eta (f) \bigr) (x,y)
=  
2\pi \sum_{mj} 
\frac{e^{-imx} \overline{f_m (y)} }
{\left| R'(y_{|m|j}) \right|}\, 
\delta(y,y_{|m|j})
\  ,
\label{eq:eta-ftilde1}
\ee
where the Dirac delta-distributions in (\ref{eq:eta-ftilde1-int}) and
(\ref{eq:eta-ftilde1}) are respectively those on $\BbbR$
and~$S^1$. 
While (\ref{eq:eta-formal}) remains formal, the right-hand sides of 
(\ref{eq:eta-ftilde1-int}) and
(\ref{eq:eta-ftilde1}) are well-defined distributions on smooth
functions on~$\mathcal{C}$. We now adopt 
(\ref{eq:eta-ftilde1}) (or equivalently (\ref{eq:eta-ftilde1-int})) 
as the 
definition of our~$\eta$ and proceed to show that this $\eta$
satisfies the rigging map axioms. 

For $f, \, g \in \Phi$, 
(\ref{eq:eta-ftilde1}) yields
\be
\eta(f) [g]
= 
{(2\pi)}^2 
\sum_{mj}
\frac{
\overline{ {f}_m (y_{|m| j})} 
{g}_m (y_{|m| j}) }
{\left| R'(y_{|m| j}) \right|}
\ . 
\label{eq:eta-ftilde-gtilde1}
\ee
From (\ref{eq:eta-ftilde-gtilde1}) it is evident that $\eta$ is real
and positive, and also that the image of $\eta$ is nontrivial and
finite dimensional. From (\ref{eq:U(t)}) it follows that vectors in
the image of $\eta$ are invariant under the dual action of~$U(t)$. We
show in appendix  
\ref{app:intertwining}
that $\eta$ intertwines with $\Aobs$ in the sense
of~(\ref{eq:eta-intertwining-matrix}). $\eta$~is thus a rigging map,
and the physical Hilbert space $\Hraq$ is the image of $\eta$ equipped
with the inner product that can be read off from (\ref{phys-ip})
and~(\ref{eq:eta-ftilde-gtilde1}). Note that no Cauchy completion is
needed since $\Hraq$ is finite dimensional.

We show in appendix 
\ref{app:irrep} 
that the representation of $\Aobs$ on
$\Hraq$ is irreducible.

As all the states in $\Hraq$ have their support in the classically
allowed region of~${\mathcal C}$, there is no tunnelling of the kind
found in \cite{AH} into the classically forbidden region of~${\mathcal
C}$. If the classically allowed region of ${\mathcal C}$ is not
connected, there is however tunnelling between all its components that
support states in~$\Hraq$.

Finally, consider the semiclassical limit. When $\hbar$ is reinstated,
equation (\ref{eq:q2=R}) becomes $\hbar^2 q^2 = R(y)$. In the limit
$\hbar\to0$, the dimension of $\Hraq$ thus asymptotes to $2/\hbar$
times the total variation of $\sqrt{R}$ over the subset of $S^1$ on
which $R$ is positive. From section
\ref{sec:classical} 
we see that this is $(2\pi\hbar)^{-1}$ times the volume of
$\Gammared$. Although our $\Gammared$ is not compact, this is the
semiclassical limit one might have expected on comparison with
geometric quantisation on compact phase spaces, such as quantisation
of angular momentum on the phase space $S^2$
\cite{woodhouse-book,sniatycki}.

\subsection{Rigging map with 
degenerate solutions to (\ref{eq:q2=R})}
\label{sec:p>1}

In this subsection we allow stationary points of $R$ among the
solutions to~(\ref{eq:q2=R}). As formulas (\ref{eq:eta-ftilde1}) and
(\ref{eq:eta-ftilde-gtilde1}) then become ill defined, some
modification is required.

We label the solutions to (\ref{eq:q2=R}) as follows. Let $p$ denote
the order of the lowest nonvanishing derivative of $R$ at a
solution. For odd~$p$, we write the solutions as~$y_{pqj}$, where the
last index enumerates the solutions with given $p$ and~$q$. For
even~$p$, we write the solutions as~$y_{p \epsilon qj}$, where
$\epsilon\in\{1,-1\}$ is the sign of the $p$th derivative and the last
index enumerates the solutions with given $p$, $\epsilon$ and~$q$.

Let $\mathcal{P}$ be the value set of the first index
of the solutions $\{y_{p qj}\}$ and $\{y_{p \epsilon qj}\}$. 
We assume that $\mathcal{P}$ has the following property: 
\begin{itemize}
\item[]
\textit{If $p\in\mathcal{P}$, then
$\mathcal{P}$ contains no factors of 
$p$ smaller than~$p/2$.} 
\end{itemize}
The case of subsection \ref{sec:p=1} is recovered for 
$\mathcal{P}=\{1\}$. 

For each odd $p\in\mathcal{P}$, we now define the map $\eta_p: \Phi
\to \Phi^*$ by
\be
\bigl(\eta_p ( f ) \bigr) (x,y)
=  
2\pi \sum_{mj}
\frac{e^{-imx} \overline{{f}_m (y)}}
{{\left| R^{(p)}(y_{p|m|j}) \right|}^{1/p}}\, 
\delta(y,y_{p|m|j})
\  . 
\label{eq:eta-p-ftilde1}
\ee
Similarly, for each even $p\in\mathcal{P}$ and 
each $\epsilon\in\{1,-1\}$ for which
solutions to (\ref{eq:q2=R}) exist, we define the map
$\eta_{p\epsilon}: \Phi \to \Phi^*$ by
\be
\bigl(\eta_{p\epsilon} ( f ) \bigr) (x,y)
=  
2\pi \sum_{mj}
\frac{e^{-imx} \overline{{f}_m (y)}}
{{\left| R^{(p)}(y_{p \epsilon |m|j}) \right|}^{1/p}}\, 
\delta(y,y_{p \epsilon |m|j})
\  . 
\label{eq:eta-peps-ftilde1}
\ee
When $\mathcal{P}=\{1\}$, the only map defined by
these formulas 
is $\eta_1$ from (\ref{eq:eta-p-ftilde1}) with $p=1$, and this map is
identical to that in~(\ref{eq:eta-ftilde1}): We thus recover the
results of subsection~\ref{sec:p=1}. When $\mathcal{P}\ne\{1\}$, the
maps (\ref{eq:eta-p-ftilde1}) and (\ref{eq:eta-peps-ftilde1}) with
$p>1$ receive contributions from precisely those solutions to
(\ref{eq:q2=R}) for which the corresponding terms in
(\ref{eq:eta-ftilde1}) diverge. We can therefore think of the maps
(\ref{eq:eta-p-ftilde1}) and (\ref{eq:eta-peps-ftilde1}) with $p>1$ as
appropriately renormalised versions of the respective ill-defined
terms in (\ref{eq:eta-ftilde1}). We show in appendix
\ref{app:intertwining} that the coefficients in
(\ref{eq:eta-p-ftilde1}) and (\ref{eq:eta-peps-ftilde1}) are fixed by
the requirement that the maps have the intertwining
property~(\ref{eq:eta-intertwining-matrix}). 

If $f , \, g\in \Phi$, 
(\ref{eq:eta-p-ftilde1}) and (\ref{eq:eta-peps-ftilde1}) 
give 
\begin{subequations}
\label{eq:eta-+-ftilde-gtilde1}
\be
\eta_p (f) [ g ]
&=& 
{(2\pi)}^2 
\sum_{mj} 
\frac{
\overline{ {f}_m (y_{p|m| j})} 
{g}_m (y_{p|m| j}) }
{{\left| R^{(p)}(y_{p|m|j}) \right|}^{1/p}}
\ , 
\label{eq:eta-p-ftilde-gtilde1}
\\
\eta_{p\epsilon} 
(f) [ g ]
&=& 
{(2\pi)}^2 
\sum_{mj} 
\frac{
\overline{ {f}_m (y_{p \epsilon |m| j})} 
{g}_m (y_{p \epsilon |m| j}) }
{{\left| R^{(p)}(y_{p \epsilon |m|j}) \right|}^{1/p}}
\ . 
\label{eq:eta-peps-ftilde-gtilde1}
\ee
\end{subequations}
From (\ref{eq:eta-+-ftilde-gtilde1}) it is seen that each $\eta_p$ and
$\eta_{p\epsilon}$ has a finite-dimensional, nontrivial image and
satisfies the rigging map axioms, with the possible exception of the
intertwining property~(\ref{eq:eta-intertwining-matrix}). We show in
appendix
\ref{app:intertwining}
that each $\eta_p$ and $\eta_{p\epsilon}$
satisfies also the intertwining property and hence provides
a rigging map. Each of the images of these maps 
provides therefore a
RAQ physical
Hilbert space, denoted respectively by 
$\mathcal{H}_\mathrm{RAQ}^{p}$ and
$\mathcal{H}_\mathrm{RAQ}^{p\epsilon}$, 
with the inner product given by
(\ref{phys-ip}) and~(\ref{eq:eta-+-ftilde-gtilde1}). 
As all the spaces are finite dimensional, 
no Cauchy completion is needed. 

As the images of any two of the rigging maps have trivial intersection
in~$\Phi^*$, we can regard 
$\mathcal{H}_\mathrm{RAQ}^{p}$ and
$\mathcal{H}_\mathrm{RAQ}^{p\epsilon}$ 
as superselection
sectors in the `total' RAQ Hilbert space 
\be 
\mathcal{H}_\mathrm{RAQ}^{\mathrm{tot}} 
:= 
\left( 
\bigoplus_{p\ \mathrm{odd}} \mathcal{H}_\mathrm{RAQ}^{p}
\right)
\oplus 
\left(
\bigoplus_{p\ \mathrm{even}, \ \epsilon} \mathcal{H}_\mathrm{RAQ}^{p
\epsilon}
\right)
\ . 
\ee
We show in appendix 
\ref{app:irrep}
that the
representation of $\Aobs$ on each 
$\mathcal{H}_\mathrm{RAQ}^{p}$ and
$\mathcal{H}_\mathrm{RAQ}^{p\epsilon}$ is irreducible. 
This means that
there are no further superselection sectors in 
$\mathcal{H}_\mathrm{RAQ}^{\mathrm{tot}}$. 

There is again no tunnelling into the classically forbidden region of
${\mathcal C}$, but within each $\mathcal{H}_\mathrm{RAQ}^{p}$ and
$\mathcal{H}_\mathrm{RAQ}^{p\epsilon}$ there can be tunnelling between
the connected components of the classically allowed region
of~${\mathcal C}$.

In the semiclassical limit, the dimension of
$\mathcal{H}_\mathrm{RAQ}^{1}$ asymptotes to $(2\pi\hbar)^{-1}$ times
the volume of $\Gammared$, while the dimension of the orthogonal
complement of $\mathcal{H}_\mathrm{RAQ}^{1}$ in
$\mathcal{H}_\mathrm{RAQ}^{\mathrm{tot}}$ remains bounded. In this
sense, 
the semiclassical limit in $\mathcal{H}_\mathrm{RAQ}^{\mathrm{tot}}$
comes entirely from the superselection sector 
$\mathcal{H}_\mathrm{RAQ}^{1}$.

\section{Concluding remarks}
\label{sec:discussion}

In this paper we have studied refined algebraic quantisation (RAQ) of
Boulware's version of the Ashtekar-Horowitz model. Although the system
did not allow a rigging map to be defined in terms of an absolutely
convergent integral of matrix elements over the gauge group, the
formal group averaging expressions nevertheless suggested a rigging
map candidate, and we showed that for generic potential functions this
candidate is a rigging map and the resulting representation of the RAQ
observable algebra $\Aobs$ on the physical Hilbert space $\Hraq$ is
irreducible. For certain special potentials the rigging map candidate
contained formally divergent terms, but a renormalisation of these
terms yielded a genuine rigging map, and in this case the
representation of $\Aobs$ on $\Hraq$ decomposed into
superselection sectors.  The dimension of $\Hraq$ was in all cases
finite and bore the expected semiclassical relation to the volume of
the reduced phase space. The only superselection sector that remained
significant in the semiclassical limit was the one whose rigging map
required no renormalisation.

The system exhibits a striking connection between the singular subsets
in the reduced phase space $\Gammared$ and the superselection sectors
in the quantum theory. Because of the periodicity of the coordinate
$x$ on the unreduced configuration space ${\mathcal C} \simeq T^2$,
the conjugate momentum $p_x$ gets quantised in integer values. For
generic potentials, these integer values entirely miss the singular,
measure zero subsets of~$\Gammared$, and in this case the quantum
theory has no superselection sectors. However, when the potential is
such that one or more of the quantised values of $p_x$ hit some of the
singular subsets of~$\Gammared$, superselection sectors arise in the
quantum theory.

Although the compactness of ${\mathcal C}$ simplified some
aspects of the analysis, the compactness is as such not essential: The
results remain qualitatively similar if the $y$-direction
is unwrapped to the real axis, provided the range of $y$ in which $R$
takes positive values remains bounded. What is essential is the
periodicity in the $x$-direction. As seen in
appendix~\ref{app:Gammared}, it is the $x$-periodicity that in the
classical theory renders the volume of $\Gammared$ finite and creates
the singular subsets; in the quantum theory, it is the associated
discreteness of $p_x$ that makes the physical Hilbert space
finite dimensional and allows the isolated stationary points of the
potential to make nonzero contributions to the rigging map. If $x$
takes values in~$\BbbR$, these phenomena do not arise. The reduced
phase space has then infinite volume and no singularities, the
physical Hilbert space is infinite dimensional and the stationary
points of $R$ make a vanishing contribution to the rigging map.

The divergences in the rigging map appear to be related to the rate of
divergence in the formal group averaging integral. It might be
possible to investigate this issue in a more precise setting in
systems where the variable $y\in S^1$ is replaced by a variable that
takes values on a higher-dimensional space, say $T^n$ with $n>1$. 
For
$n\ge3$ and an $R$ whose only stationary points are nondegenerate, 
it should then be possible to make the convergence of the averaging so
strong that the uniqueness theorem of Giulini and Marolf 
\cite{GM2} applies and implies in particular 
that there are no superselection sectors. Modifying the stationary
point structure of $R$ should then offer a range of options for
weakening the sense of convergence and creating superselection
sectors.

One would like to understand whether the connection between classical
singularities and quantum superselection sectors extends from our
specific system to more general classes of constrained systems. On the
classical side, the structure of the reduced phase space at the
singularities can be described by the methods of singular
Marsden-Weinstein reduction~\cite{SjaLer}. On the quantum side, the
methods of RAQ can be regarded as a version of Rieffel
induction~\cite{Rieffel}, which has been argued
\cite{Lands,Lands-Bialo} to provide a natural quantum
counterpart of the Marsden-Weinstein reduction. The basic language for
discussing this connection appears therefore to be in place. It is
likely that results in this direction would involve criteria on the
RAQ test space: In the Rieffel induction language, such criteria can
be described in terms of spectral continuity~\cite{Lands-Bialo}.

Our quantum theory appears physically reasonable, and when there are
no superselection sectors, the theory can be regarded as a
specification of operators on the physical Hilbert space constructed
already in~\cite{Boul}. As the system was originally introduced in
\cite{Boul} 
as a simplified version of the Ashtekar-Horowitz (AH) model~\cite{AH},
one might expect our methods to produce a physically reasonable
quantisation also for the AH model. In the AH model the variables $x$
and $y$ are interpreted as respectively the azimuthal and longitudinal
angle on~$S^2$, so that $y$ has period $2\pi$ but $0 \le x \le \pi$,
where the limits correspond to coordinate singularities at the north
and south poles. The classical constraint can be promoted into a
self-adjoint operator by introducing suitable boundary conditions at
$x=0$ and $x=\pi$~\cite{Boul}, and we can then proceed essentially as
in the Boulware system, recovering a finite-dimensional physical
Hilbert space. Whether this quantisation is physically reasonable
seems now to hinge on one one's viewpoint on the classical system. On
the one hand, if $x=0$ and $x=\pi$ are regarded as classically
excluded, the reduced phase space has infinite volume, and one would
then expect a quantum theory with an infinite-dimensional Hilbert
space \cite{Tate,Ash2}. This viewpoint is adopted in the context of
algebraic quantisation in \cite{Tate,Ash2}. On the other hand, the
incompleteness of the Hamiltonian vector field of the constraint
suggests that one might want to interpret the classical theory at
$x=0$ and $x=\pi$ in terms of reflective boundary conditions of some
sort, for example as suggested by free motion at constant longitude on
the round sphere. This viewpoint gives the reduced phase space finite
volume, which leads one to expect a finite-dimensional Hilbert space
in the quantum theory. Our quantisation methods are thus compatible
with the latter classical viewpoint. A~refined algebraic quantisation
of the AH model that would be compatible with the former classical
viewpoint remains an intriguing open problem.

\section*{Acknowledgements}
We thank Don Marolf for drawing our attention to 
refined algebraic quantisation of the 
Ashtekar-Horowitz model 
and 
John Barrett, 
Joel Feinstein, 
Nico Giulini 
and 
Jo\~ao Martins
for helpful discussions. 
We thank the referee for raising the connection to singular
Marsden-Weinstein reduction. 
J.L. thanks Arundhati Dasgupta, Jack Gegenberg and Viqar
Husain for discussions and 
the opportunity to present this work at the April 2005
Quantum Gravity Workshop at the University of New Brunswick. 
A.M. was
supported by a CONACYT (Mexico) Postgraduate Fellowship and an ORS
Award to the University of Nottingham.

\appendix

\section{Appendix: 
$\Gammared$}
\label{app:Gammared}

In this appendix we verify the properties of $\Gammared$ stated in the
main text.


Each orbit generated by the constraint $C$ on the constraint
hypersurface $\barGamma$ has constant $y$ and~$p_x$. We consider first
the subset of $\barGamma$ where $p_x \ne 0$, which is always nonempty,
and then include the subset (if nonempty) where $p_x=0$.

\subsection{$p_x\ne0$}

Let $I \subset S^1$ be an open interval in which $R$ takes positive
values. The corresponding two subsets of $\barGamma$ are 
${\mathcal N} := 
\bigl\{ 
\bigl(x,y, \sqrt{R(y)} \, , p_y \bigr) 
\mid 
x \in S^1
, \, 
y\in I 
, \, 
p_y \in \BbbR 
\bigr\}$
and a similar set with a minus sign in front of the square root. We
consider~${\mathcal N}$; the situation for the other set is similar.

The orbits that $C$
generates in ${\mathcal N}$ have constant~$y$, 
and they satisfy $\dot{x} \ne0$ and 
$\dot{p}_y / \dot{x} = \tfrac12 R'/\sqrt{R}$. 
If $x$ were not periodic, we could
choose from each of the gauge orbits 
a unique point by the condition $x=0$ and
hence represent the projection of ${\mathcal N}$ to $\Gammared$ as
${\mathcal M}:=
\bigl\{ 
\bigl(0,y,\sqrt{R(y)}\, , p_y \bigr) 
\mid 
y\in I 
, \, 
p_y \in \BbbR 
\bigr\} 
\simeq
I \times \BbbR$, with the symplectic form 
$\Omegared = dp_y \wedge dy$. As 
$x$ is periodic, however, 
the projection of ${\mathcal N}$ to
$\Gammared$ is not ${\mathcal M}$ 
but instead the quotient space ${\mathcal M}/\BbbZ$, where the
$\BbbZ$-action is 
\begin{equation}
(y,p_y) \overset{n}\longmapsto 
\bigl( y,p_y + \pi n R'(y)/\sqrt{R(y)}\, \bigr) 
\ , \ n\in\BbbZ 
\ . 
\label{eq:Z-action}
\end{equation}

The structure of ${\mathcal M}/\BbbZ$ now depends on whether $I$
contains stationary points of~$R$.


\subsubsection{No stationary points}

If $I$ does not contain stationary points of~$R$,
the $\BbbZ$-action (\ref{eq:Z-action}) 
is properly discontinuous and 
${\mathcal M}/\BbbZ$ is a symplectic 
manifold with topology $I \times S^1$. 
We can introduce on ${\mathcal M}$ the adapted coordinates 
$(w,\theta)$ by $w = \sqrt{R(y)}$ and 
$\theta = 2 p_y \sqrt{R(y)} / R'(y)$, 
in which the symplectic form reads 
$\Omegared = d\theta \wedge dw$ and the 
$\BbbZ$-action (\ref{eq:Z-action}) takes the form 
\begin{equation}
(w,\theta) \overset{n}\longmapsto 
(w,\theta + 2\pi n ) 
\ , \ n\in\BbbZ 
\ , 
\label{eq:Z-action-ad}
\end{equation}
so that $\theta$ becomes periodic with period $2\pi$
in ${\mathcal M}/\BbbZ$. 
The symplectic volume of ${\mathcal M}/\BbbZ$ 
is finite and equal to $\pi \int_I |R'(y)| / \sqrt{R(y)} \, dy$, 
or $2\pi$ times the total variation of $\sqrt{R}$ over~$I$.

\subsubsection{Stationary points}

If $I$ contains stationary points of~$R$, we may assume without loss
of generality that $y_0 \in I$ is the only such stationary point. We
write $I = (y_-,y_+)$, $I_+ := (y_0, y_+)$ and $I_- := (y_-, y_0)$, so
that $I$ is the disjoint union of $I_+$, $I_-$ and~$\{y_0\}$. The
$\BbbZ$-action (\ref{eq:Z-action}) is then properly discontinuous for
$y\in I_+$ and $y\in I_-$ but not at $y=y_0$. ${\mathcal M}/\BbbZ$
consists thus of two open cylinders, coming respectively from $I_+$
and $I_-$ and each being a symplectic manifold, joined together by a
line at $y=y_0$. ${\mathcal M}/\BbbZ$ is clearly connected.
It is not Hausdorff, since points on the line at $y=y_0$ do not have
disjoint neighbourhoods. 

We shall show that ${\mathcal M}/\BbbZ$ is not a manifold. We first
construct a subset of ${\mathcal M}/\BbbZ$ that is homeomorphic to
$\BbbR^2$ and then show that the properties of this subset prevent
${\mathcal M}/\BbbZ$ from being a manifold.

To begin, let ${\mathcal M}_+ := 
\bigl\{ 
\bigl(0,y,\sqrt{R(y)}\, , p_y \bigr) \in {\mathcal M}
\mid 
y\in I_+
\bigr\}$ and $q := \bigl(0,y_0,\sqrt{R(y_0)}\, , 0 \bigr) 
\in 
{\mathcal M}$. 
The $\BbbZ$-action (\ref{eq:Z-action}) restricts to
${\mathcal M}_+$ and to 
${\mathcal M}_+ \cup \{q\}$. 

We introduce in ${\mathcal M}_+$ the adapted coordinates $(w, \theta)$
as above. Writing $w_0 := \sqrt{R(y_0)}$ and $w_+ := \sqrt{R(y_+)}$,
we then introduce in ${\mathcal M}_+/\BbbZ$ the coordinates $(u,v)$ by
$u = \sqrt{|w-w_0|} \cos(\theta)$ and $v = \sqrt{|w-w_0|}
\sin(\theta)$, where $0<u^2+v^2< | w_+ - w_0 |$.

Let ${\mathcal S}\simeq \BbbR^2$ denote the space obtained by adding to 
${\mathcal M}_+/\BbbZ$ in the 
chart $(u,v)$ the point $u=0=v$. 
Given in ${\mathcal M}_+/\BbbZ$ a sequence of points
that converges to the point $u=0=v$ in ${\mathcal S}$, 
we can choose in ${\mathcal M}_+$ a sequence of pre-images 
that converges to $q$ in ${\mathcal M}_+ \cup \{q\}$. This 
shows that 
${\mathcal S} \simeq \bigl({\mathcal M}_+ \cup \{q\}\bigr)/\BbbZ$, 
where the homeomorphism holds in the sense of topological manifolds. 

Two side remarks are in order. 
First, although we here 
only need ${\mathcal S}$ to be a topological manifold, 
we note that the symplectic form on 
${\mathcal M}_+/\BbbZ$ continues into a symplectic form on 
${\mathcal S}$ in the differentiable structure determined by the chart 
$(u,v)$: We have 
$\Omegared = 2 \, \mathrm{sign}
\bigl(w_+ - w_0 \bigr) 
dv \wedge du$. 
Second, although the above construction made a 
specific choice for the point~$q$, 
a similar construction can be given if $q$ 
is replaced by any point in ${\mathcal M}$ at $y=y_0$. 

Suppose now that ${\mathcal M}/\BbbZ$ is a manifold. If so, it has to
be two-dimensional. Let $\bar{q} \in {\mathcal M}/\BbbZ$ be the
projection of the point~$q$, and let $U\simeq \BbbR^2$ be a
neighbourhood of $\bar{q}$ in ${\mathcal M}/\BbbZ$. As $\bar{q} \in
{\mathcal S}$, $U \cap {\mathcal S}$ is nonempty and open
in~${\mathcal S}$. As ${\mathcal S}$ is a two-manifold, there exists a
set $V \subset U \cap {\mathcal S}$ such that $\bar{q} \in V$ and
$V\simeq\BbbR^2$. Since $V\subset U$, $V$ is open as a subset of~$U$,
and since $U$ is open in ${\mathcal M}/\BbbZ$, $V$~is open in
${\mathcal M}/\BbbZ$. But this is a contradiction since every
neighbourhood of $\bar{q}$ in ${\mathcal M}/\BbbZ$ contains points
that are not in~$V$.\footnote{We are grateful to Nico Giulini for
pointing out that in a similar argument given in \cite{lou-ma-torus},
p.~318, the space denoted therein by $\overline{\mathcal M}_t$ should
be replaced by the space in which the tilded holonomy parameters take
arbitrary values, and this space is not the closure of~${\mathcal
M}_t$.}

\subsection{$p_x=0$}

Suppose that $y_\mathrm{b}\in S^1$ such that $R(y_\mathrm{b})=0$. 

By our assumptions about~$R$, 
the inverse function theorem \cite{lang-real} implies that  
there exists an open interval~$J$, symmetric about~$0$, 
in which the equation 
$p_x^2 = R(y)$ can be solved for $y$ as $y=F(p_x)$, where 
$F: J\to\BbbR$ is a smooth even function, $F(0) = y_\mathrm{b}$ 
and the only stationary point of $F$ is~$0$. 
The corresponding subset of $\barGamma$ is ${\mathcal Q} := 
\bigl\{ 
\bigl(x,F(p_x), p_x , p_y \bigr) 
\mid 
x \in S^1
, \, 
p_x \in J 
, \, 
p_y \in \BbbR 
\bigr\}$.
The orbits that $C$
generates in ${\mathcal Q}$ have constant~$p_x$, 
and they satisfy $\dot{p}_y \ne0$ and
$\dot{x} / \dot{p}_y = F'(p_x)$. 
Adopting the gauge $p_y=0$, 
we find as above that the projection of ${\mathcal Q}$ to 
$\Gammared$ can be represented as the set  
$\bigl\{ 
\bigl(x,F(p_x), p_x , 0 \bigr) 
\mid 
x \in S^1
, \, 
p_x \in J 
\bigr\}
\simeq S^1 \times \BbbR$ 
and the symplectic form reads $\Omegared = dp_x \wedge dx$.

\section{Appendix: 
The rigging maps are intertwiners}
\label{app:intertwining}

In this appendix we show that the rigging maps defined in the main
text have the intertwining
property~(\ref{eq:eta-intertwining-matrix}).
We work under the assumptions of 
subsection~\ref{sec:p>1}, recovering in the 
special case $\mathcal{P}=\{1\}$ the situation 
of subsection~\ref{sec:p=1}. 

Let $A\in\Aobs$. Let $m$ and $n$ be fixed integers and let 
$f, \, g \in
\Phi$ such that 
$f(x,y) = e^{imx}f_m(y)$ 
and 
$g(x,y) = e^{inx}g_n(y)$. 
As $U(t)$ is unitary and commutes with~$A^\dag$,
we have 
${\bigl(U(-t)f, A^\dag g\bigr)}_{\mathrm{aux}}
= 
{\bigl(f, U(t) A^\dag g\bigr)}_{\mathrm{aux}}
= 
{\bigl(f, A^\dag U(t) g\bigr)}_{\mathrm{aux}}
=
{\bigl(A f, U(t) g\bigr)}_{\mathrm{aux}}$. 
Using (\ref{eq:U(t)}) in the leftmost and rightmost expressions and
performing the integration over $x$ in the inner products gives 
\be
\int\! dy\, 
e^{it[R(y) - m^2] } 
\overline{f_m(y)} \bigl(A^\dag g\bigr)_m(y) 
=
\int\! dy\,
e^{it[R(y)-n^2]} 
\overline{\bigl(A f\bigr)_n(y)} g_n (y)
\ ,
\label{eq:iidentity}
\ee
where for any 
$h\in\Phi$ and $B\in\Aobs$ 
we have introduced the notation 
$\bigl(Bh\bigr)(x,y) =: 
\sum_k e^{ikx} \bigl(Bh\bigr)_k(y)$. 

On each side of~(\ref{eq:iidentity}), we break the integral over $y\in
S^1$ into a sum over integrals over open intervals $\{I_\alpha\}$
whose end-points are adjacent stationary 
points of~$R$. Let $R_\alpha$ be the
restriction of $R$ to~$I_\alpha$, and let $R_\alpha^{-1}$ be the
inverse of~$R_\alpha$. Changing the integration variable in each of
the integrals on the left-hand side to 
$s:= R_\alpha(y) - m^2$ and on the right-hand side 
to $s:= R_\alpha(y) - n^2$, we obtain 
\begin{equation}
\int\! ds\, 
e^{its } 
\sum_\alpha
\left[ 
\frac{\overline{f_m} \, \bigl(A^\dag g\bigr)_m}
{|R'|}
\right]
\bigl( R_\alpha^{-1} (s+m^2) \bigr)
=
\int\! ds\,
e^{its} 
\sum_\alpha
\left[ 
\frac{\overline{\bigl(A f\bigr)_n} \, g_n}
{|R'|}
\right]
\bigl( R_\alpha^{-1} (s+n^2) \bigr)
\ , 
\label{eq:ifourier}
\end{equation}
where for given $s$ the sum on the left-hand side (right-hand side) is
over the values of $\alpha$ for which $s+m^2$ (respectively $s+n^2$)
is in the image of $R_\alpha$. The integrand on the left-hand side
(right-hand side) is not defined at the stationary values of $R-m^2$
(respectively $R-n^2$), which are finitely many, but it is continuous
in $s$ elsewhere and defines an $L^1$-function of the variable
$s\in\BbbR$. 

Now, regarded as a function of~$t\in\BbbR$, each side of
(\ref{eq:ifourier}) is the Fourier transform of an $L^1$-function. 
(\ref{eq:ifourier})~therefore 
implies the $L^1$ equality 
\begin{equation}
\sum_\alpha
\left[ 
\frac{\overline{f_m} \, \bigl(A^\dag g\bigr)_m}
{|R'|}
\right]
\bigl( R_\alpha^{-1} (s+m^2) \bigr)
=
\sum_\alpha
\left[ 
\frac{\overline{\bigl(A f\bigr)_n} \, g_n}
{|R'|}
\right]
\bigl( R_\alpha^{-1} (s+n^2) \bigr)
\ , 
\label{eq:pointwise}
\end{equation}
and the continuity
observations above imply that the equality in (\ref{eq:pointwise})
holds pointwise in $s$ except at the stationary values of
$R-m^2$ and $R-n^2$.

Consider the right-hand side of 
(\ref{eq:pointwise}) as a function of~$s$. 
In a sufficiently small punctured neighbourhood of $s=0$, this
function is a sum of contributions 
in which a local inverse of $R$ takes $s+n^2$ close to some 
$y_{p|n|j}$ and contributions 
in which two local inverses of $R$ take $s+n^2$ 
close to some~$y_{p\epsilon
|n|j}$. 
An elementary analysis shows that the 
contribution from near 
$y_{p|n|j}$ has the asymptotic small $s$ expansion 
\begin{equation}
\frac{{(p!)}^{1/p}}{p s} 
\left(
\frac{\overline{\bigl(A f\bigr)_n (y_{p|n|j}) } \, g_n (y_{p|n|j})}
{{\left| R^{(p)}(y_{p|n|j}) \right|}^{1/p}}
s^{1/p}
+
\sum_{k=2}^\infty 
a_k 
{\bigl(s^{1/p}\bigr)}^{k}
\right)
\ ,
\label{eq:deg:oddp}
\end{equation}
where the fractional power 
$s^{1/p}$ stands for the branch that has the same sign as~$s$. 
Similarly, the contribution from near 
$y_{p \epsilon |n|j}$
has the asymptotic small $s$ expansion 
\begin{equation}
2 \theta(\epsilon s) 
\frac{{(p!)}^{1/p}}{p |s|} 
\left( 
\frac{\overline{\bigl(A f\bigr)_n (y_{p\epsilon |n|j}) } 
\, g_n (y_{p \epsilon|n|j})}
{{\left| R^{(p)}(y_{p \epsilon |n|j}) \right|}^{1/p}}
{|s|}^{1/p} 
+ 
\sum_{k=3}^\infty b_k 
{|s|}^{k/p} 
\right)
\ , 
\label{eq:deg:evenp}
\end{equation}
where $\theta$ is the Heaviside function. We have suppressed in
(\ref{eq:deg:oddp}) and (\ref{eq:deg:evenp}) the various indices on
the coefficients $a_k$ and~$b_k$. The factor $2$ in
(\ref{eq:deg:evenp}) arises because there are two contributing local
inverses, and the sum in (\ref{eq:deg:evenp}) lacks a $k=2$ term
because the $k=2$ terms from the two contributing local inverses
cancel.

Similar considerations apply to the left-hand side
of~(\ref{eq:pointwise}).

By the property of the index set 
$\mathcal{P}$ stated in 
subsection~\ref{sec:p>1}, 
the leading contribution from each $p$ 
in the small $s$ expansion of (\ref{eq:pointwise}) 
has a power distinct from that 
of sub-leading contributions from any higher~$p$. 
Equating the coefficients order by order 
thus shows that for each odd~$p$ 
\be
\sum_{j}
\frac{
\overline{ {f}_m (y_{p|m| j})} 
\bigl(A^\dag g\bigr)_m (y_{p|m| j}) }
{{\left| R^{(p)}(y_{p|m| j}) \right|}^{1/p}}
= 
\sum_{j}
\frac{
\overline{ \bigl(A f\bigr)_n (y_{p |n| j})} 
g_n (y_{p |n| j}) }
{{\left| R^{(p)}(y_{p|n| j}) \right|}^{1/p}}
\ , 
\label{eq:oddp-rr}
\ee
while for each even $p$ and each $\epsilon\in\{1,-1\}$ 
\be
\sum_{j}
\frac{
\overline{ {f}_m (y_{p\epsilon|m| j})} 
\bigl(A^\dag g\bigr)_m (y_{p\epsilon|m| j}) }
{{\left| R^{(p)}(y_{p\epsilon|m| j}) \right|}^{1/p}}
= 
\sum_{j}
\frac{
\overline{ \bigl(A f\bigr)_n (y_{p\epsilon|n| j})} 
g_n (y_{p\epsilon|n| j}) }
{{\left| R^{(p)}(y_{p\epsilon|n| j}) \right|}^{1/p}}
\ . 
\label{eq:evenp-rr}
\ee
In terms of the maps $\eta_p$ and 
$\eta_{p\epsilon}$~(\ref{eq:eta-+-ftilde-gtilde1}), 
(\ref{eq:oddp-rr}) and (\ref{eq:evenp-rr}) read 
\begin{subequations}
\label{eq:intert-dd-fg}
\be
\eta_p(f) [A^\dag g]
&=& 
\eta_p(A f) [g]
\ , 
\\
\eta_{p\epsilon}(f) [A^\dag g]
&=& 
\eta_{p\epsilon}(A f) [g]
\ . 
\ee
\end{subequations}
By linearity, these arguments leading to (\ref{eq:intert-dd-fg})
continue to hold when $f$ and $g$ are replaced by arbitrary
vectors in~$\Phi$. Hence each $\eta_p$ and $\eta_{p\epsilon}$ has the
intertwining property~(\ref{eq:eta-intertwining-matrix}).

Finally, we note that the assumption about the index set $\mathcal{P}$
can be replaced by weaker assumptions that involve
also~$\epsilon$. For example, if only one sign of $\epsilon$ is known
to occur, it suffices to assume that the even and odd subsets of
$\mathcal{P}$ individually have the property stated in
subsection~\ref{sec:p>1}.

\section{Appendix: Representation of $\Aobs$} 
\label{app:irrep}

In this appendix we show that the representation of 
$\Aobs$ on each of the Hilbert spaces 
$\mathcal{H}_\mathrm{RAQ}^{p}$ and
$\mathcal{H}_\mathrm{RAQ}^{p\epsilon}$ 
of subsection \ref{sec:p>1} is irreducible. The irreducibility on the 
Hilbert space of subsection \ref{sec:p=1} follows as the special case 
$\mathcal{P}=\{1\}$. 

We discuss the cases of odd and even $p$ separately. 

\subsection{$\mathcal{H}_\mathrm{RAQ}^{p}$} 

Fix an odd $p\in\mathcal{P}$. 
To unclutter the notation, 
we will suppress $p$ in most of the formulas. 

We first construct a set of tailored observables. 

For each $y_{qj}$ and $y_{rk}$ (where the index $p$ is suppressed) 
we define 
a function $h_{qj;rk}$ from a neighbourhood of 
$y_{qj}$ to a neighbourhood of $y_{rk}$ by the formula 
\be
h_{qj;rk}(y) := 
R^{-1}_{rk}\left( R(y)-q^2+r^2 \right)
\ , 
\label{eq:h-def}
\ee
where $R^{-1}_{rk}$ is the inverse of the restriction of $R$ to a
neighbourhood of~$y_{rk}$. Raising both sides of the equation
$R(y)-q^2 = R(h) - r^2$ to power $1/p$ and applying the implicit
function theorem \cite{lang-real} shows that $h_{qj;rk}$ is
well-defined and smooth, and we can choose the domains to be pairwise
disjoint and such that $\bigl(h_{qj;rk}\bigr)^{-1} = h_{rk;qj}$.

For each $y_{qj}$, 
we choose a smooth function $\rho_{qj}$ on~$S^1$, 
such that $\rho_{qj}(y_{qj}) =1$ 
and the support of $\rho_{qj}$ is contained 
in the domain of $h_{qj;rk}$ for all~$y_{rk}$. 

We now define on $\Phi$ the operators $A_{mj;nk}$ (where the index $p$
is suppressed) such that if $f\in\Phi$, $f(x,y) = \sum_l
e^{ilx}f_l(y)$, then
\be
\bigl(A_{mj;nk} f \bigr) (x,y) 
= e^{imx}\rho_{|m|j}(y)
f_n\left(h_{|m|j;|n|k}(y)\right)
\ . 
\label{eq:A-action}
\ee
In words, $A_{mj;nk}$ first annihilates from $f$ all components except 
the one whose $x$-dependence is~$e^{inx}$, 
then 
modifies in this component the function of $y$ 
to the zero function everywhere except near $y=y_{|n|k}$, 
and finally 
maps to a vector whose $x$-dependence is~$e^{imx}$, 
with $y$-dependence nonzero only near $y=y_{|m|j}$. 

A direct computation shows that each $A_{mj;nk}$ commutes
with~$U(t)$. The adjoint of $A_{mj;nk}$ acts on $\Phi$ as
\be
\Bigl( {\bigl(A_{mj;nk}\bigr)}^\dag f \Bigr) (x,y) 
= e^{inx}
\overline{\rho_{|m|j}\Bigl(\bigl(h_{|n|k;|m|j}(y)\bigr) \Bigr)}
f_m\left(h_{|n|k;|m|j}(y)\right)
\ , 
\label{eq:Adag-action}
\ee
and comparison of 
(\ref{eq:A-action}) and (\ref{eq:Adag-action})
shows that also the adjoint
commutes with~$U(t)$. Each $A_{mj;nk}$ is 
therefore in~$\Aobs$. 

With this preparation, we can prove: 
\begin{proposition}
\label{proposition:irred}
Let $V\subset \mathcal{H}_\mathrm{RAQ}^{p}$ 
be a linear subspace invariant under~$\Aobs$, 
$V \ne \{0\}$. Then
$V = \mathcal{H}_\mathrm{RAQ}^{p}$. 
\end{proposition}

\myproof
Let $v\in V$, $v\ne0$. Let $u\in \Phi$ such that $v=\eta(u)$. 
We write $u(x,y) = \sum_l e^{ilx} u_l(y)$. From 
(\ref{eq:eta-p-ftilde-gtilde1}) it follows that there exist $n$ and
$k$ such that $u_n(y_{p|n|k}) \ne0$. 

For each 
$m$ and $j$ such that $y_{p|m|j}$ exists, 
we now define $w^{mj} := 
A_{mj;nk} u$. It follows that $\eta (w^{mj}) \in V$, 
and from the construction of $A_{mj;nk}$ 
we see that for every $f\in\Phi$, 
\be
\eta (w^{mj})[f] = {(2\pi)}^2 
\frac{\overline{u_n(y_{p|n|k})}}
{{\left| R^{(p)}(y_{p|m|j}) \right|}^{1/p}}
\, 
f_m (y_{p|m|j})
\ .
\label{eq:wmj-on-f}
\ee
Comparison of (\ref{eq:wmj-on-f}) and 
(\ref{eq:eta-p-ftilde-gtilde1}) shows that  
the set $\{\eta(w^{mj})\}$ spans $\mathcal{H}_\mathrm{RAQ}^{p}$. 
$\blacksquare$

\subsection{$\mathcal{H}_\mathrm{RAQ}^{p\epsilon}$} 

Fix an even $p\in\mathcal{P}$, and 
fix $\epsilon$ so that 
solutions $y_{p\epsilon |m|j}$ exist. 

No restriction of $R$ to a neighbourhood of $y_{p\epsilon |m|j}$ now
has an inverse. However, we can give a meaning to (\ref{eq:h-def}) by
raising both sides of the equation $\epsilon \bigl[R(y)-q^2\bigr] =
\epsilon \bigl[R(h) - r^2\bigr]$ to power~$1/p$, with the branches
chosen so that (say) both sides are increasing, and applying the
implicit function theorem. After this, the arguments go through as for
odd~$p$.


\end{document}